\documentclass[12pt]{article}
\usepackage{amssymb}
\begin{document}
\baselineskip=18pt
\pagenumbering{arabic}
\parskip1.5em
\newcommand{\ee}{{{\eta}\over {2}}}
\newcommand{\beq}{\begin{equation}}
\newcommand{\eeq}{\end{equation}}
\newcommand{\beqa}{\begin{eqnarray}}
\newcommand{\eeqa}{\end{eqnarray}}
\newcommand{\beqan}{\begin{eqnarray*}}
\newcommand{\eeqan}{\end{eqnarray*}}
\newcommand{\half}{{{1}\over{2}}}
\newcommand{\ihalf}{{{i}\over{2}}}
\newcommand{\quar}{{{1}\over{4}}}
\newcommand{\la}{\lambda}
\newcommand{\si}{\sigma}
\newcommand{\Si}{\Sigma}
\newcommand{\tf}{\Theta}
\newcommand{\3}{{\ss}}
\newcommand{\bra}{\langle 0|}
\newcommand{\ket}{|0\rangle}
\newcommand{\id}{{1}\hspace{-0.3em}{\rm{I}}}
\newcommand{\tn}[1]{T^{#1}}
\newcommand{\tr}{\bigtriangleup}
\newcommand{\trb}{\bar{\bigtriangleup}}
\def\b{\beta}
\def\d{\delta}
\def\g{\gamma}
\def\a{\alpha}
\def\s{\sigma}
\def\t{\tau}
\def\l{\lambda}
\def\e{\epsilon}
\def\r{\rho}
\def\d{\delta}
\def\wid{\widehat}
\def\ds{\displaystyle}
\def\be{\begin{equation}}
\def\ee{\end{equation}}
\def\beq{\begin{eqnarray}}
\def\eeq{\end{eqnarray}}
\def\ov{\overline}
\def\om{\omega}
\thispagestyle{empty}
\begin{flushright}BN--TH--2000--05\\
SFB 288 Preprint No. 467\\
\end{flushright}
\vskip2.5em
\begin{center}
{\Large{\bf The Drinfel'd twisted $XYZ$ model}}\vskip1.5em
T.-D. Albert$^{1}$\hspace{3em}H. Boos$^{2\,\natural}$\hspace{3em}R. Flume$^{1}$\hspace{3em}R.H. Poghossian$^{1\,\flat}$\hspace{3em}K. Ruhlig$^{1}$\\\vskip3em
$^{1}${\sl Physikalisches Institut der Universit\"at Bonn}\\
{\sl Nu{\ss}allee 12, D--53115 Bonn, Germany}\\\vskip1.5em
$^{2}${\sl Institut f\"ur Theoretische Physik}\\
{\sl Freie Universit\"at Berlin, Arnimallee 14, D--14195 Berlin, Germany}
\end{center}
\vskip2em
\begin{abstract}
\noindent We construct a factorizing Drinfel'd twist for a face type model equivalent to the $XYZ$ model. Completely symmetric expressions for the operators of the monodromy matrix are obtained.
\end{abstract}
$^{\natural}$ {\small{on leave of absence from the Institute of High Energy Physics, Protvino, Russia}}\\
$^{\flat}$ {\small{on leave of absence from Yerevan Physics Institute, Armenia}}\\
{\small{
e-mail: t-albert@th.physik.uni-bonn.de\\
\hspace*{1.3cm}boos@physik.fu-berlin.de\\
\hspace*{1.3cm}flume@th.physik.uni-bonn.de\\
\hspace*{1.3cm}poghos@th.physik.uni-bonn.de\\
\hspace*{1.3cm}ruhlig@th.physik.uni-bonn.de}}
\vfill\eject\setcounter{page}{1}

\section{Introduction}
Several integrable quantum spin chain models within the range of the
algebraic Bethe ansatz method have a distinguished basis of states, which minimalizes quantum effects. That is, the quasiparticle creation and
annihilation operators in this basis have an appearance devoid of
polarization clouds. For the $XXX$ and $XXZ$ models with underlying group
$sl(2)$ the bases in question were found by Maillet and Sanchez de
Santos \cite{ms} through the construction of a generalized Drinfeld twist \cite{drinfeld}.\\ 
The ensuing representation of the quantum monodromy matrices coincides, as noted by Terras \cite{terras}, for the case of the rational $XXX$ model with the representation provided by Sklyanin's functional Bethe ansatz method \cite{sklyanin1}.
An obvious generalization of Sklyanin's method (substituting 
polynomials in the spectral parameter by polynomials in the exponential of the spectral parameter) leads us to the conclusion that an analogous coincidence holds true for the trigonometric $XXZ$ model.\par\noindent
The purpose of the present communication is to report on the generalization of the above results to the elliptic $sl(2)$-$XYZ$ model. For this sake we will make use of Baxter's map of the $XYZ$ model onto an ice type model \cite{baxter1} (which is akin to a  cyclic solid on solid model \cite{pearce}). This brings us  formally near to the $XXX$ and $XXZ$ models and allows
us to  use the technique developed in \cite{abfr} for the handling of the analogous problem in the $sl(n)$-$XXX$ model. The plan of the paper is as
follows : Section 2 provides a short survey of the $XYZ$ model and its reformulation as an ice-type model. Section 3 deals with the factorizing twists and the computation of the operator valued entries of the monodromy matrix. Section 4 contains the conclusions. The appendix is devoted to proving some relation needed in Sect.3.   
\section{$XYZ$ model and its relation to ice-type models}
In the framework of the Algebraic Bethe Ansatz \cite{fadtakh} the $XYZ$-model is determined by the elliptic solution of the Yang-Baxter equation 
\beqa
R_{12}(\la_1-\la_2)R_{13}(\la_1-\la_3)R_{23}(\la_2-\la_3)=R_{23}(\la_2-\la_3)R_{13}(\la_1-\la_3)R_{12}(\la_1-\la_2)
\eeqa
with
\beqa
R^{xyz}(\la-\mu)=\left(
\begin{array}{c c c c}a&0&0&d\\0&b&c&0\\0&c&b&0\\d&0&0&a\end{array}
\right)\label{rxyz}
\eeqa
where
\beqa
a(\la-\mu)&=&{{\Theta(2\eta)\Theta(\la-\mu)}\over{\Theta(0)\Theta(\lambda-\mu+2\eta)}}\nonumber\\
b(\la-\mu)&=&{{\Theta(2\eta)H(\la-\mu)}\over{\Theta(0)H(\lambda-\mu+2\eta)}}\nonumber\\
c(\la-\mu)&=&{{H(2\eta)\Theta(\la-\mu)}\over{\Theta(0)H(\lambda-\mu+2\eta)}}\nonumber\\
d(\la-\mu)&=&{{H(2\eta)H(\la-\mu)}\over{\Theta(0)\Theta(\lambda-\mu+2\eta)}}
\eeqa
with the notation $H(u)=\vartheta_1\left({{u}\over{2K}},q\right),\;\Theta(u)=\vartheta_4\left({{u}\over{2K}},q\right)$ and $\vartheta_4=\sum_{m\in\mathbb{Z}}(-1)^n q^{n^2}e^{2\pi i nz},\;\vartheta_1(z,q)=-i q^{{1\over 4}}e^{i\pi z}\theta_4(z+\tau/2,q)$ are the standard theta-functions of a single complex variable.\footnote{For a concise introduction into the theory of theta and elliptic functions we refer the reader to the appendix of \cite{fadtakh}, whose conventions we shall use throughout our work.}\\
The somewhat different parametrization as compared to \cite{fadtakh} is due to the normalization in order to achieve unitarity for the R-matrix.\\\\
The monodromy matrix  $T(\la,\{\la_i\})$ (generalized to the inhomogeneous chain \cite{evi}, \cite{sklytak}) is given as the ordered product of Lax operators $L_i(\la-\la_i)=R_{0i}(\la-\la_i)$
\beqa
T(\la,\left\{\la_i\right\})=L_N(\la-z_N)\ldots L_2(\la-z_2)L_1(\la-z_1)=\left(\begin{array}{c c }A(\la,\left\{\la_i\right\})&B(\la,\left\{\la_i\right\})\\C(\la,\left\{\la_i\right\})&D(\la,\left\{\la_i\right\}) \end{array}
\right)\label{monodromy}\;.
\eeqa
The presence of the Boltzmann weight $d$ in Eq. (\ref{rxyz}) reflects the non-conservation of spin, which is responsible for the absence of a local vacuum for the Lax operator associated with the above R-matrix.\\
To circumvent the problems arising from the eight vertex nature, we use the vertex--face map established by Baxter \cite{baxter1} to obtain a $XXZ$ type (six vertex) R-matrix by exploiting the relation
\beqa
R^{xyz}(\la-\mu)\phi_{l,l'}\otimes z_{m',l'}=\sum_{m}w(m,m'|l,l')\phi_{m,m'}\otimes z_{m,l} \label{vfm}
\eeqa
valid for all integers $l,l',m,m'$ such that $|l-l'|=|m-m'|=1$ and the summation on the r.h.s. is over integers s.t.  $|m-m'|=1,\;\;|m-l|=1$.\\
The two dimensional vectors $\phi,\;z$ are given by
\beqa
\phi_{l,l+1}&=&X(s_l+\mu);\;\;\;z_{l+1,l}=X(s_l+\la)\nonumber\\
\phi_{l+1,l}&=&X(t_{l+1}-\mu);\;\;\;z_{l-1,l}=X(t_l-\la)
\eeqa
with $X(u)={H(u)\choose \Theta(u)}$ and the abbrevation $s_l=s+2\eta l$, where $s,t$ are arbitrary complex parameters.\\
Relation (\ref{vfm}) provides a change of basis in which the Lax operator has a local vacuum independent of the spectral parameter $\la$. There exist a whole family of bases labeled by the integer $l$ and the parameters $s,t$, which achieve this goal.\\ 
The six weights $w(m,m'|l,l')$ are represented graphically in Figure 1.\\
\vspace*{4.5cm}
\begin{picture}(800, 100)(0,100)\thicklines\setlength{\unitlength}{0.355mm}\linethickness{1pt}
\put(50,50){\line(0,1){100}}
\put(100,100){\line(-1,0){100}}
\put(25,20){$a_l=w(l-1,l|l-2,l-1)$}\put(20,70){$l-1$}\put(70,70){$l$}\put(20,130){$l-2$}\put(70,130){$l-1$}
\put(25,100){\line(-2,1){10}}\put(25,100){\line(-2,-1){10}}
\put(85,100){\line(-2,1){10}}\put(85,100){\line(-2,-1){10}}
\put(50,75){\line(1,-2){5}}\put(50,75){\line(-1,-2){5}}
\put(50,125){\line(1,-2){5}}\put(50,125){\line(-1,-2){5}}

\put(200,50){\line(0,1){100}}
\put(250,100){\line(-1,0){100}}
\put(175,20){$b_l=w(l+1,l|l,l-1)$}\put(170,70){$l+1$}\put(220,70){$l$}\put(170,130){$l$}\put(220,130){$l-1$}
\put(175,100){\line(-2,1){10}}\put(175,100){\line(-2,-1){10}}
\put(235,100){\line(-2,1){10}}\put(235,100){\line(-2,-1){10}}
\put(200,65){\line(-1,2){5}}\put(200,65){\line(1,2){5}}
\put(200,115){\line(-1,2){5}}\put(200,115){\line(1,2){5}}

\put(350,50){\line(0,1){100}}
\put(400,100){\line(-1,0){100}}
\put(325,20){$c_l=w(l+1,l|l,l+1)$}\put(320,70){$l+1$}\put(370,70){$l$}\put(320,130){$l$}\put(370,130){$l+1$}
\put(325,100){\line(-2,1){10}}\put(325,100){\line(-2,-1){10}}
\put(375,100){\line(2,-1){10}}\put(375,100){\line(2,1){10}}
\put(350,65){\line(-1,2){5}}\put(350,65){\line(1,2){5}}
\put(350,125){\line(1,-2){5}}\put(350,125){\line(-1,-2){5}}
\end{picture}\\

\begin{picture}(800, 100)(17,10)\thicklines
\setlength{\unitlength}{0.355mm}\linethickness{1pt}
\put(50,50){\line(0,1){100}}
\put(100,100){\line(-1,0){100}}
\put(25,20){$a'_l=w(l+1,l|l+2,l+1)$}\put(20,70){$l+1$}\put(70,70){$l$}\put(20,130){$l+2$}\put(70,130){$l+1$}
\put(15,100){\line(2,-1){10}}\put(15,100){\line(2,1){10}}
\put(75,100){\line(2,-1){10}}\put(75,100){\line(2,1){10}}
\put(50,65){\line(-1,2){5}}\put(50,65){\line(1,2){5}}
\put(50,125){\line(-1,2){5}}\put(50,125){\line(1,2){5}}

\put(200,50){\line(0,1){100}}
\put(250,100){\line(-1,0){100}}
\put(175,20){$b'_l=w(l-1,l|l,l+1)$}\put(170,70){$l-1$}\put(220,70){$l$}\put(170,130){$l$}\put(220,130){$l+1$}
\put(165,100){\line(2,-1){10}}\put(165,100){\line(2,1){10}}
\put(225,100){\line(2,-1){10}}\put(225,100){\line(2,1){10}}
\put(200,75){\line(1,-2){5}}\put(200,75){\line(-1,-2){5}}
\put(200,135){\line(1,-2){5}}\put(200,135){\line(-1,-2){5}}

\put(350,50){\line(0,1){100}}
\put(400,100){\line(-1,0){100}}
\put(325,20){$c'_l=w(l-1,l|l,l-1)$}\put(320,70){$l-1$}\put(370,70){$l$}\put(320,130){$l$}\put(370,130){$l-1$}
\put(315,100){\line(2,-1){10}}\put(315,100){\line(2,1){10}}
\put(385,100){\line(-2,1){10}}\put(385,100){\line(-2,-1){10}}
\put(350,75){\line(1,-2){5}}\put(350,75){\line(-1,-2){5}}
\put(350,125){\line(-1,2){5}}\put(350,125){\line(1,2){5}}
\end{picture}\\
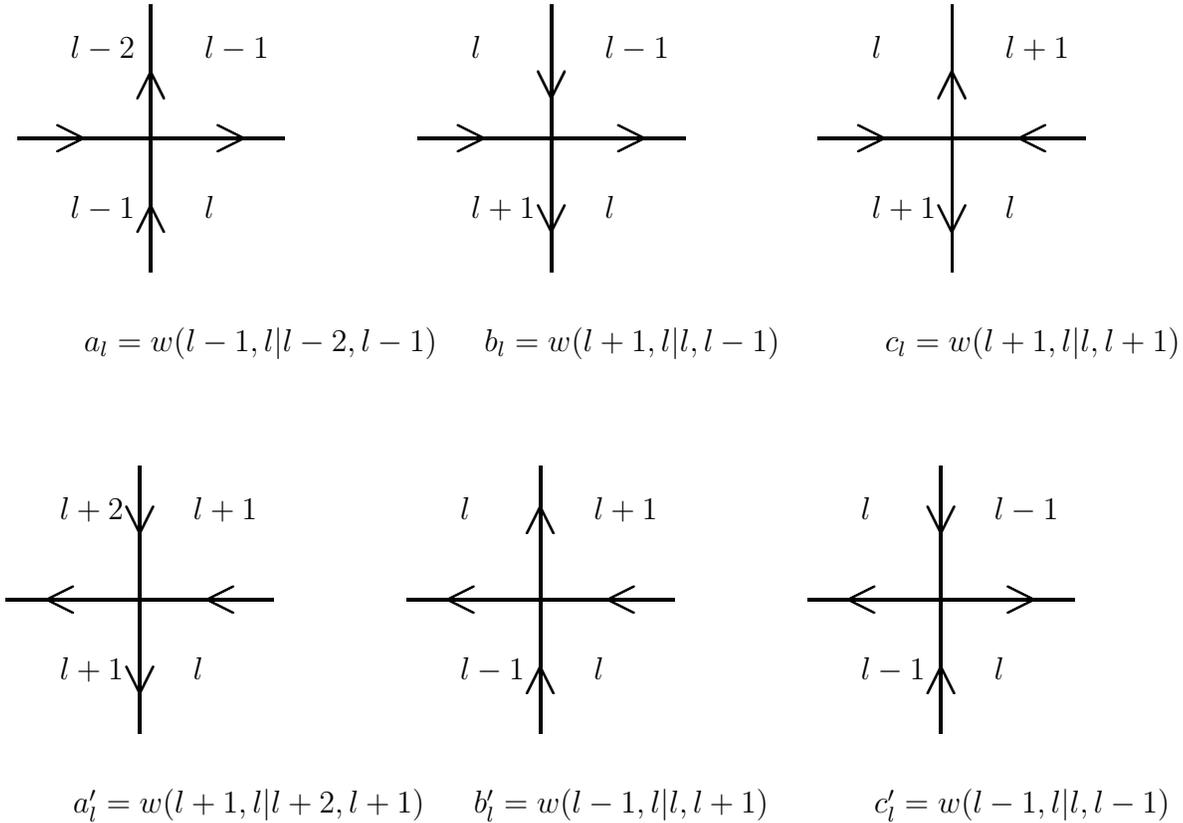
{\small{FIG. 1 The six resulting Boltzmann weights.}}\\\\
The Boltzmann weights are subject to the star-triangle relation
\beq
\sum_{m}w(l_2,l_3|l_1,m)w(l_3,l_4|m,l_5)w(m,l_5|l_1,l_6)=\sum_{m}w(l_3,l_4|l_2,m)w(l_2,m|l_1,l_6)w(m,l_4|l_6,l_5)\;.
\eeq
They are parametrized by $\left(h(u)=\Theta(0)H(u)\Theta(u);\;\;\omega_l=({{s+t}\over{2}}+2\eta l-K)\right)$:
\beqa
a_l&=&a'_l=1;\nonumber\\&&\nonumber\\
b_l&=&{{h(\la)h(\omega_{l-1})}\over{h(\la+2\eta)h(\omega_l)}};\;\;b'_l={{h(\la)h(\omega_{l+1})}\over{h(\la+2\eta)h(\omega_l)}}\nonumber\\&&\nonumber\\
c_l&=&{{h(2\eta)h(\omega_l-\la)}\over{h(\la+2\eta)h(\omega_l)}};\;\;c'_l={{h(2\eta)h(\omega_l+\la)}\over{h(\la+2\eta)h(\omega_l)}}\label{weights}
\eeqa
These weights can be arranged into a matrix
\beqa
R_{12}(l)=\left(
\begin{array}{c c c c}a_l&0&0&0\\0&b_l&c_l&0\\0&c'_l&b'_l&0\\0&0&0&a'_l\end{array}
\right)\label{Rmatrix}
\eeqa
which fulfills the modified Yang-Baxter equation \cite{gn}, \cite{felder}
\beqa
R_{12}(l-\sigma_3)R_{13}(l)R_{23}(l-\sigma_1)=R_{23}(l)R_{13}(l-\sigma_2)R_{12}(l)\label{ybemod}\;.
\eeqa
The monodromy matrix related to this modified Yang-baxter equation is \footnote{We use the convention $X_{0,1\ldots N}$ to denote an operator $X$ represented in space $0$ with entries in the tensorproduct of spaces $1,\ldots,N$.}
\beq
T_{0,1\ldots N}(l)=R_{0N}(l-\sigma_1-\ldots -\sigma_{N-1})\ldots R_{02}(l-\sigma_1)R_{01}(l)\label{T(l)}
\eeq
where $0$ denotes the horizontal auxiliary space (with the asssociated spectral parameter $\la_0$), $1,\ldots,N$ label  the vertical quantum spaces which span the physical Hilbertspace ${\cal{H}}_N$ (with associated local inhomogeneities $\left\{\la_i\right\}$), and $\sigma_i$ equals $\pm 1$ depending on whether the arrow in the $i$-th space is up or down (right/left for the horizontal space). It also sets our convention to associate the integer in the right lower corner of the graphical representation with the operator (cf. Figure 2).\\\vspace*{1.5cm}
\setlength{\unitlength}{0.2mm}\linethickness{1pt}
\begin{picture}(800, 150)(-50,50)
\put(150,70){\line(1,0){300}}
\put(210,0){\line(0,1){140}}
\put(270,0){\line(0,1){140}}
\put(330,0){\line(0,1){140}}
\put(390,0){\line(0,1){140}}
\put(500,70){...........}
\put(590,70){\line(1,0){150}}
\put(620,0){\line(0,1){140}}
\put(690,0){\line(0,1){140}}
\put(760,65){$j$}
\put(-60,65){$\bigg(T(l)_{0,1\ldots N}\bigg)_{ij}=$}
\put(137,65){$i$}
\put(200,155){$N$}
\put(260,155){$N-1$}
\put(325,155){$N-2$}
\put(385,155){$N-3$}
\put(615,155){${2}$}
\put(685,155){$1$}
\put(720,10){$l$}
\put(630,10){$l-\sigma_1$}
\put(520,10){$l-\sigma_1-\sigma_2$}
\put(100,10){$l-\sum_{i=1}^N\sigma_i$}
\end{picture}\\
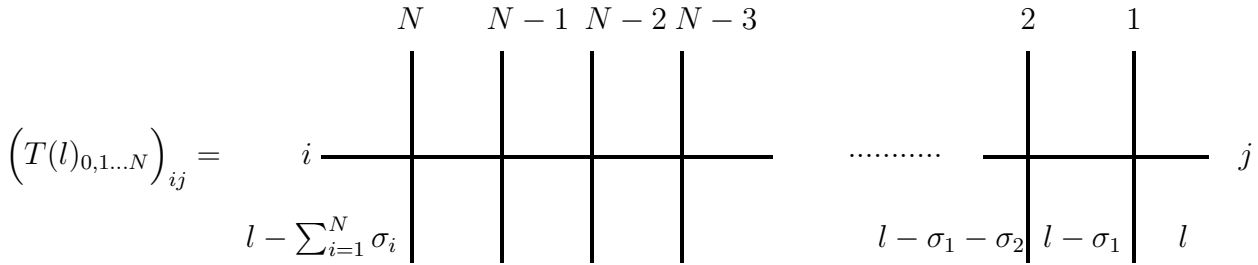
{\small{FIG. 2 Elements of the monodromy matrix}}\\\\
From (\ref{ybemod}) follows the equation for the monodromy matrices
\beq
R_{00'}(l-\sigma_1-\ldots-\sigma_N)T_{0}(l)\otimes T_{0'}(l-\sigma_{0})=T_{0'}(l)\otimes T_{0}(l-\sigma_{0'})R_{00'}(l)\label{ybe}
\eeq
and can be represented graphically as in Figure 3.\\
\vspace*{4cm}
\setlength{\unitlength}{0.1mm}
\begin{picture}(1000, 150)(10,210)\thicklines\linethickness{1pt}
\put(150,30){\line(1,0){310}}\put(150,30){\line(-1,0){75}}
\put(200,-140){\line(0,1){230}}\put(230,90){\oval(60,60)[tl]}\put(230,120){\line(1,0){230}}
\put(350,-140){\framebox(10,410)}
\put(400,-30){$l$}
\put(550,35){=}
\put(750,-140){\framebox(10,410)}
\put(650,120){\line(1,0){400}}
\put(650,20){\line(1,0){230}}\put(880,50){\oval(60,60)[br]}\put(910,50){\line(0,1){210}}\put(940,10){$l$}
\put(200,-185){$0'$}
\put(30,15){$0$}
\put(900,270){$0'$}
\put(1100,110){$0$}
\put(350,-185){${\cal{H}}_N$}
\put(750,-185){${\cal{H}}_N$}
\end{picture}\\
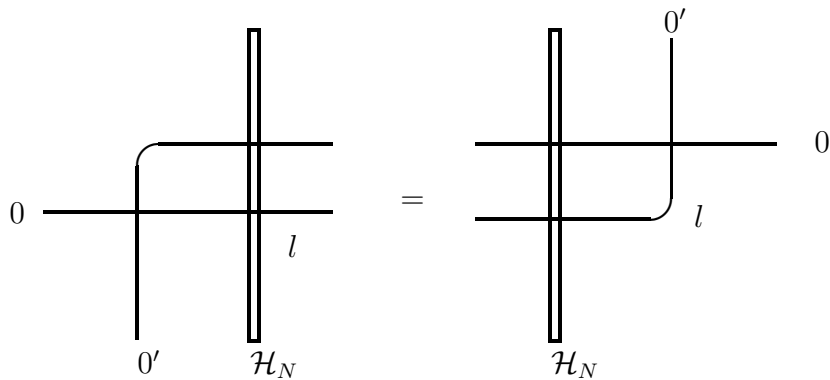
{\small{FIG. 3 The Yang--Baxter equation for the monodromy matrix}}\\\\
The unitarity relation $R_{21}R_{12}=\id$ can be checked directly by using the following relation
\beqa
&&h(u+x)h(u-x)h(v+y)h(v-y)-h(u+y)h(u-y)h(v+x)h(v-x)\nonumber\\
&&=h(x-y)h(u-y)h(v+x)h(v+x)\label{h}
\eeqa
which is a consequence of the equality $H(u,k)\Theta(u,k)=H((1+k)u,2\sqrt{k}/(1+k))$ with $k$ the nome of the elliptic function (Eq. 15.10.20 of \cite{baxbook}), and the analogue of (\ref{h}) for $H(u)$ (Eq. 15.3.10 of \cite{baxbook}).\\\\
The representation of eigenvalues of the transfer matrix (\ref{monodromy}) through these of (\ref{T(l)}) is explained in reference \cite{fadtakh} (One has however to keep in mind that our model (\ref{T(l)}) differs from that of \cite{fadtakh} by an additional change of basis in the quantum space).\\
We will concentrate in what follows on the computation of a factorizing $F$-matrix for the monodromy matrix (\ref{T(l)}) for an arbitrary fixed value of $l$.
\section{The $F$ basis}
The factorizing $F$-matrix for two sites defined by the relation $F_{21}R_{12}=F_{12}$ is
\beqa
F_{12}=\left(
\begin{array}{c c c c}1&0&0&0\\0&1&0&0\\0&c'&b'&0\\0&0&0&1\end{array}
\right)_{[12]}\;.\label{Fmatrix12}
\eeqa
The proof of the factorization property amounts to checking the same relations as those in the proof of the unitarity of the R-matrix above.\par\noindent
The factorizing $F$-matrix for $N$ sites ($N$ quantum spaces) turns out to be given by formally the same expression as found in \cite{abfr} for the $XXX$ model\footnote{We thank Frank G\"ohmann for communicating this representation of $F_{1\ldots N}$, which is slightly simpler than that quoted in \cite{abfr}.}
\beqa
F_{1\ldots N}(l)&=& \sum_{{{\alpha}}\in {\mathbb{Z}}_2^N}
P_{{\alpha}} R_{1\ldots N}^{\sigma_{{\alpha}}}(l)(z_{1},\ldots,z_{N})\nonumber\\
P_{{\alpha}}&=&\prod_{i=1}^N P_{i}^{\alpha_i}\label{Fmatrix}
\eeqa
where $P_{i}^{\alpha_i}$ projects on the ${\alpha_i}$-th component in the $i$-th space and the permutation $\sigma_{\alpha}$ is uniquely determined through the conditions 
\beqa
\alpha_{\sigma_{\alpha}(i+1)}&\geq&\alpha_{\sigma_{\alpha}(i)}\quad \mbox{if}\quad\sigma_{\alpha}(i+1)>\sigma_{\alpha}(i)\nonumber\\
\alpha_{\sigma_{\alpha}(i+1)}&>&\alpha_{\sigma_{\alpha}(i)}\quad \mbox{if}\quad\sigma_{\alpha}(i+1)<\sigma_{\alpha}(i)\;.
\eeqa
An algorithm for finding $\sigma_{\alpha}$ for a given $\alpha$ is described in the appendix.\\
The modification of the Yang-Baxter equation (\ref{ybemod}) enforces a particular rule for the handling of the integer valued parameter $l$ in the formation of the intertwining matrix $R^{\sigma}(l)$
(related to the permutation $\sigma$), which can be read off from the modified composition law
\beqa
R^{\sigma\sigma_{i}}(l)&=&R_{\sigma(i),\sigma(i+1)}(\tilde{l}_i)R^{\sigma}(l)\nonumber\\
\tilde{l}_i&=&l-\sigma_{\sigma(1)}-\ldots -\sigma_{\sigma(i-1)}\label{complaw}
\eeqa
where $\sigma_{i}$ is the transposition of $i,i+1$, and $\sigma$ an arbitrary permutation.\\
$R^{\sigma}(l)$ has the intertwining property 
\beqan
R^{\sigma}(l)T_{0,1\ldots N}(l)=T_{0,\sigma(1)\ldots \sigma(N)}(l)R^{\sigma}(l-\sigma_0)\;.
\eeqan
The matrix $F_{1\ldots N}(l)$ satisfies the factorizing equation 
\beqa
R^{\sigma}_{1\ldots N}(l)=F_{\sigma(1\ldots N)}^{-1}(l)F_{1\ldots N}(l)\;.\label{ffe}
\eeqa
A proof of the latter equation can be found in \cite{abfr} -- the modification of the composition law going along with the parameter $l$ being rather immaterial. An alternative argument emphasizing the geometric meaning of (\ref{ffe}) (loosely speaking a pure gauge field representation of $R^{\sigma}$ on the symmetric group ${\cal{S}}_N$ as base space) may be sketched as follows:\\
It is sufficient to prove Eq. (\ref{ffe}) for arbitrary elementary transpositions $\sigma=\sigma_i$;\,$\sigma_i:\;(i,i+1)\rightarrow(i+1,i)$,
\beqa
R^{\sigma_i}_{1\ldots N}=F^{-1}_{\sigma_i(1\ldots N)}F_{1\ldots N}\;.\label{pg}
\eeqa
A special solution $\hat{F}^{i}$ of (\ref{pg}) is the two site matrix $F_{i\,i+1}$, Eq. (\ref{Fmatrix12}), embedded into the N-fold tensor product 
\beqan
\hat{F}^{i}=\id\otimes \ldots\otimes F_{i\,i+1}\otimes\ldots\otimes\id
\eeqan
and the most general solution of (\ref{pg}) is given by
\beqa
F_{1\ldots N}=X(1,\ldots,\{i,i+1\},\ldots, N)\hat{F}^{i}
\eeqa
where $X(1,\ldots,\{i,i+1\},\ldots, N)$ denotes a non-singular matrix in the quantum space $\otimes^N{\mathbb{C}}^2$ which is symmetric with respect to an exchange of the space labels $i$ and $i+1$ and of the respective local inhomogeneity parameters. So we note that the proof of Eq. (\ref{ffe}) is complete if we can show that 
\beqa
F_{1\ldots N}\left({\hat{F}}^{i}\right)^{-1}=X_{(1,\ldots N)}^{(i)}\label{ffx}
\eeqa
is symmetric in the labels $i$ and $i+1$. For this purpose it is convenient to consider the permutation group element $\sigma_{\alpha}$ of Eq. (\ref{ffx}) as being decomposed into products of elementary transpositions, which also leads to a factorization of $R^{\sigma_{\alpha}}$ into a product of elementary (two site) matrices $R^{\sigma_{x}}$, and to subdivide the sum $\sum_{\alpha}$ in Eq. (\ref{Fmatrix}) into terms with the transposition $\sigma_i$ and into terms without. It is furthermore possible to arrange the order of the transpositions such that $R^{\sigma_i}$ does occur at most once and then on the very right end of the respective product of matrices $R^{\sigma_{x}}$. Schematically we  obtain 
\beqa
\sum_{\alpha}P_{\alpha}R^{\sigma_{\alpha}}&=&\sum_{\alpha'}P_{\alpha'}R^{\sigma_{\alpha'}}R^{\sigma_i}+ \sum_{\alpha''}P_{\alpha''}R^{\sigma_{\alpha''}}  \label{sumdecomp}\;.
\eeqa
Each factor $P_{\alpha'}R^{\sigma_{\alpha'}}$ in the first sum has a counterpart $P_{\alpha''}R^{\sigma_{\alpha''}}$ in the second sum related by an interchange of the labels $i$ and $i+1$.\\
Keeping this in mind one verifies straightforwardly by inspection that the multiplication of Eq. (\ref{sumdecomp}) by
\beqan
\hat{F}_{\sigma_i}^{-1}&=&\left(\sum_{2\geq\alpha_{i}>\alpha_{i+1}}\id_{i,i+1} P_{i}^{\alpha_i}P_{i+1}^{\alpha_{i+1}}+\sum_{2\geq\alpha_{i+1}\geq\alpha_{i}}R^{\sigma_i}P_{i}^{\alpha_i}P_{i+1}^{\alpha_{i+1}}\right)\,\times\,diag\{1,b^{-1}(\la_j,\la_i),b^{-1}(\la_i,\la_j),1\}\\&&
\eeqan
from the right side produces an expression which is symmetric in $i$ and $i+1$. \par\noindent
The operators of the monodromy matrix (\ref{T(l)}) in the $F$ basis are obtained by using a  recursion relation, which enables one to express the monodromy matrix for $N$ sites in terms of that for $N-1$ sites. Starting from the definition ${\tilde{T}}_{0,1\ldots N}(l)=F_{1\ldots N}(l)T_{0,1\ldots N}(l)F_{1\ldots N}^{-1}(l-\sigma_0)$ we use the identity $F_{1\ldots N}(l)=F_{2\ldots N}(l-{{1+\sigma_1}\over{2}})F_{1,2\ldots N}(l)$  with $F_{1,2\ldots N}(l)=\left(P_1^1+P_1^2 T_{1,23\ldots N}\right)$ (for the proof of this identity we refer the reader to the appendix)
\beqan
&&
{\tilde{T}}_{0,1\ldots N}(l)=F_{1\ldots N}(l)T_{0,1\ldots N}(l)F_{1\ldots N}^{-1}(l-\sigma_0)\nonumber\\&&\nonumber\\
&=&F_{2\ldots N}(l-{{1+\sigma_1}\over{2}})F_{1,2\ldots N}(l)T_{0,1\ldots N}(l)F_{1,2\ldots N}^{-1}(l-\sigma_0)F_{2\ldots N}^{-1}(l-\sigma_0-{{1+\sigma_1}\over{2}})\;.\nonumber\\
\eeqan
Splitting $T_{0,12\ldots N}(l)$ into $T_{0,2\ldots N}(l-\sigma_1)R_{01}(l)$ we arrive at 
\beqan
&&
{\tilde{T}}_{0,1\ldots N}(l)\nonumber\\
&=&F_{2\ldots N}(l-{{1+\sigma_1}\over{2}})F_{1,2\ldots N}(l)\,F^{-1}_{2\ldots N}(l-\sigma_1)F_{2\ldots N}(l-\sigma_1)\,T_{0,2\ldots N}(l-\sigma_1)R_{01}(l)\,\times\nonumber\\
&&\;\;\;\;\times\;F_{2\ldots N}^{-1}(l-\sigma_0-\sigma_1)F_{2\ldots N}(l-\sigma_0-\sigma_1)\,F_{1,2\ldots N}^{-1}(l-\sigma_0)F_{2\ldots N}^{-1}(l-\sigma_0-{{1+\sigma_1}\over{2}})\nonumber\\&&\nonumber\\
&=&{\tilde{F}}_{1,2\ldots N}(l){\tilde{T}}_{0,2\ldots N}(l-\sigma_1)R_{01}(l){\tilde{F}}_{1,2\ldots N}^{-1}(l-\sigma_0)\nonumber\\&&\nonumber\\
\eeqan
Inserting the explicit form of
 \beqan
{\tilde{F}}_{1,2\ldots N}(l)&=&F_{2\ldots N}\left(l-{{1+\sigma_1}\over{2}}\right)F_{1,2\ldots N}(l)F_{2\ldots N}^{-1}(l-\sigma_1)=\left(\begin{array}{c c }\id&0\\{\tilde{C}}^1_{2\ldots N}(l)&{\tilde{D}}^1_{2\ldots N}(l)\end{array}
\right)_{[1]}
\eeqan
we obtain the final expression
\beqa
{\tilde{T}}_{0,1\ldots N}(l)
=\left(\begin{array}{c c }\id&0\\{\tilde{C}}^1_{2\ldots N}(l)&{\tilde{D}}^1_{2\ldots N}(l)\end{array}
\right)_{[1]}{\tilde{T}}_{0,2\ldots N}(l-\sigma_1)R_{01}(l)\left(\begin{array}{c c }\id&0\\{\tilde{C}}^1_{2 \ldots N}(l-\sigma_0)&{\tilde{D}}^1_{2 \ldots N}(l-\sigma_0)\end{array}
\right)^{-1}_{[1]}\label{recrel}\;.\nonumber\\
\eeqa
This relation can be solved recursively starting with the  one site monodromy matrix which coincides with the Lax operator $L_i=R_{0i}$ (\ref{Rmatrix}).\par\noindent
We demonstrate the derivation and possible difficulties and their resolution in the case of the operator ${\tilde{D}}_{1\ldots N}(l)$:
\beqan
{\tilde{D}}_{1\ldots N}(l)&=&\left(\begin{array}{c c }{\tilde{D}}^0_{2\ldots N}(l-1)b'_{01}(l)&0\\X&{\tilde{D}}^1_{2\ldots N}(l){\tilde{D}}^0_{2\ldots N}(l+1)\left({\tilde{D}}^1_{2\ldots N}\right)^{-1}(l+1) \end{array}
\right)_{[1]}
\eeqan
where the lower left element $X$ is given by
\beqan
{\tilde{C}}^1_{2\ldots N}(l){\tilde{D}}^0_{2\ldots N}(l-1)b'_{01}(l)+{\tilde{D}}^1_{2\ldots N}(l){\tilde{C}}^0_{2\ldots N}(l+1)c_{01}(l)-{\tilde{D}}^0_{2\ldots N}(l){\tilde{C}}^1_{2\ldots N}(l)
\eeqan
This combination vanishes identically by virtue of the Yang--Baxter equation (\ref{ybe}), or in graphical representation:\\
\vspace*{4cm}
\setlength{\unitlength}{0.1mm}
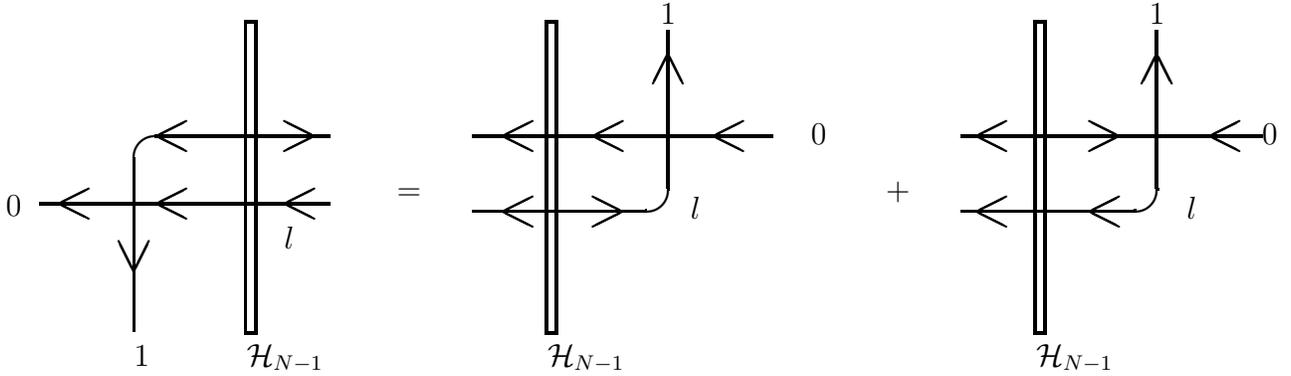
\begin{picture}(2000, 150)(10,210)\thicklines\linethickness{1pt}
\put(150,30){\line(1,0){310}}\put(150,30){\line(-1,0){75}}
\put(200,-140){\line(0,1){230}}\put(230,90){\oval(60,60)[tl]}\put(230,120){\line(1,0){230}}
\put(350,-140){\framebox(10,410)}
\put(400,-30){$l$}
\put(550,35){=}
\put(750,-140){\framebox(10,410)}
\put(650,120){\line(1,0){400}}
\put(650,20){\line(1,0){230}}\put(880,50){\oval(60,60)[br]}\put(910,50){\line(0,1){210}}\put(940,10){$l$}
\put(1200,35){+}
\put(1400,-140){\framebox(10,410)}
\put(1300,120){\line(1,0){400}}
\put(1300,20){\line(1,0){230}}\put(1530,50){\oval(60,60)[br]}\put(1560,50){\line(0,1){210}}\put(1600,10){$l$}
\put(200,-185){$1$}
\put(30,15){$0$}
\put(900,270){$1$}
\put(1100,110){$0$}
\put(350,-185){${\cal{H}}_{N-1}$}
\put(750,-185){${\cal{H}}_{N-1}$}
\put(1550,270){$1$}
\put(1700,110){$0$}
\put(1400,-185){${\cal{H}}_{N-1}$}
\put(230,30){\line(2,-1){40}}\put(230,30){\line(2,1){40}}
\put(230,120){\line(2,-1){40}}\put(230,120){\line(2,1){40}}
\put(100,30){\line(2,-1){40}}\put(100,30){\line(2,1){40}}
\put(400,30){\line(2,-1){40}}\put(400,30){\line(2,1){40}}
\put(440,120){\line(-2,1){40}}\put(440,120){\line(-2,-1){40}}
\put(200,-60){\line(-1,2){20}}\put(200,-60){\line(1,2){20}}

\put(690,20){\line(2,-1){40}}\put(690,20){\line(2,1){40}}
\put(690,120){\line(2,-1){40}}\put(690,120){\line(2,1){40}}
\put(810,120){\line(2,-1){40}}\put(810,120){\line(2,1){40}}
\put(850,20){\line(-2,1){40}}\put(850,20){\line(-2,-1){40}}
\put(970,120){\line(2,-1){40}}\put(970,120){\line(2,1){40}}
\put(910,230){\line(-1,-2){20}}\put(910,230){\line(1,-2){20}}

\put(1320,20){\line(2,-1){40}}\put(1320,20){\line(2,1){40}}
\put(1320,120){\line(2,-1){40}}\put(1320,120){\line(2,1){40}}
\put(1510,120){\line(-2,1){40}}\put(1510,120){\line(-2,-1){40}}
\put(1470,20){\line(2,-1){40}}\put(1470,20){\line(2,1){40}}
\put(1630,120){\line(2,-1){40}}\put(1630,120){\line(2,1){40}}
\put(1560,230){\line(-1,-2){20}}\put(1560,230){\line(1,-2){20}}
\end{picture}\\\\
We thus end with a diagonal matrix of the form
\beqan
{\tilde{D}}_{1\ldots N}^{0}(l)&=&\left(\begin{array}{c c }{\tilde{D}}^0_{2\ldots N}(l-1)b'_{01}(l)&0\\0&{\tilde{D}}^0_{2\ldots N}(l) \end{array}
\right)_{[1]}
\eeqan
which can now be solved iteratively
\beqan
&&\left(\begin{array}{c c }{\tilde{D}}^0_{2\ldots N}(l-1)b'_{01}(l)&0\\0&{\tilde{D}}^0_{2\ldots N}(l) \end{array}
\right)_{[1]}={\tilde{D}}_{2\ldots N}^{0}(l-{{1+\sigma_1}\over{2}})\left(\begin{array}{c c }b'_{01}(l)&0\\0&1 \end{array}
\right)_{[1]}\nonumber\\&&\nonumber\\&&\nonumber\\
&=&{\tilde{D}}_{3\ldots N}^{0}(l-{{1+\sigma_1}\over{2}}-{{1+\sigma_2}\over{2}})\left(\begin{array}{c c }b'_{02}(l-{{1+\sigma_1}\over{2}})&0\\0&1 \end{array}
\right)_{[2]}\,\otimes\,\left(\begin{array}{c c }b'_{01}(l)&0\\0&1 \end{array}
\right)_{[1]}\nonumber\\
&&\vdots\nonumber\\
&=&\otimes_{i=1}^N\left(\begin{array}{c c }b'_{0i}(l-{{1+\sigma_1}\over{2}}-{{1+\sigma_2}\over{2}}-\ldots -{{1+\sigma_{i-1}}\over{2}})&0\\0&1 \end{array}
\right)_{[i]}\;.
\eeqan\\
The computation of the operators ${\tilde{B}}_{1\ldots N}(l),\;{\tilde{C}}_{1\ldots N}(l)$ proceeds along the same lines to give
\beqan
{\tilde{B}}_{1\ldots N}(l)&=&\left(\begin{array}{c c }{\tilde{B}}^0_{2\ldots N}(l-1)b'_{01}(l)&0\\Y&{\tilde{D}}^1_{2\ldots N}(l){\tilde{B}}^0_{2\ldots N}(l+1)\left({\tilde{D}}^1_{2\ldots N}\right)^{-1}(l+1) \end{array}
\right)_{[1]}
\eeqan
with $$Y={\tilde{C}}^0_{2\ldots N}(l){\tilde{B}}^0_{2\ldots N}(l-1)b'_{01}(l)+{\tilde{D}}^1_{2\ldots N}(l){\tilde{A}}^0_{2\ldots N}(l+1)c_{01}(l)-{\tilde{D}}^1_{2\ldots N}(l){\tilde{B}}^0_{2\ldots N}(l+1)\left({\tilde{D}}^1_{2\ldots N}\right)^{-1}(l+1){\tilde{C}}^1_{2\ldots N}(l+1)$$ and for $\tilde{C}$
\beqan
{\tilde{C}}_{1\ldots N}(l)&=&\left(\begin{array}{c c }Z&{\tilde{D}}^0_{2\ldots N}(l-1)\left({\tilde{D}}^0_{2\ldots N}\right)^{-1}(l-1)c'_{01}(l)\\0&{\tilde{C}}^0_{2\ldots N}(l)\end{array}
\right)_{[1]}
\eeqan
with $$Z={\tilde{C}}^0_{2\ldots N}(l-1)-{\tilde{D}}^0_{2\ldots N}(l-1)\left({\tilde{D}}^0_{2\ldots N}\right)^{-1}(l-1){\tilde{C}}^1_{2\ldots N}(l-1)c'_{01}(l)\;.$$\par\noindent
The solution of the recursion relation finally yields the  result for the operators of the monodromy matrix $\bigg($we use a slight change in notation: $b(\la)={{h(\la)}\over{h(\la+2\eta)}}$ and denote $k=l-\sum_{i=1}^{N-1}\sigma_i$$\bigg)$:
\beqa
{\tilde{D}}_{l}(\la_0)&=&{{h(\omega_{l+1})}\over{h(\omega_{1+{{k+l-N}\over{2}}})}}\otimes_{i=1}^{N}\left(\begin{array}{c c }b(\la_{0}-\la_i)&0\\0&1\end{array}
\right)_{[i]}\nonumber\\&&\nonumber\\&&\nonumber\\
{\tilde{B}}_{l}(\la_0)&=&{{h(\omega_{l+1})}\over{h(\omega_{k})}}\sum_{i=1}^N c_{k-1}(\la_{0}-\la_i)\sigma_i^{-}\otimes_{j\neq i}^{N}\left(\begin{array}{c c }b(\la_{0}-\la_j)&0\\0&b^{-1}(\la_{j}-\la_i)\end{array}
\right)_{[j]}\nonumber\\&&\nonumber\\&&\nonumber\\
{\tilde{C}}_{l}(\la_0)&=&\sum_{i=1}^N c'_{l}(\la_{0}-\la_i)\sigma_i^{+}\otimes_{j\neq i}^{N}\left(\begin{array}{c c }b(\la_{0}-\la_j)b^{-1}(\la_i-\la_j)&0\\0&1\end{array}
\right)_{[j]}\nonumber\\\label{ops}
\eeqa
The above mentioned  basis transformation (\ref{vfm}) amounts to splitting the model into sectors with a fixed number of turned spins. To obtain the spectrum of the $XYZ$ model one uses the operators $\tilde{B}_l(\la), \tilde{C}_l(\la)$ to construct eigenvectors in the form proposed by \cite{fadtakh} (there denoted by $B_{k,l}(\la)$ etc.).\par\noindent
The remaining operator ${\tilde{A}}_{l}(\la_0)$ is obtained from the quantum determinant $det_q T(\la_0)$ \cite{kor}, \cite{sklyanin}\beqa
det_q T(\la_0)&=&{{h(\omega_{k+1})}\over{h(\omega_{l+1})}}D_{k,l}(\la_0)A_{k+1,l+1}(\la_0+2\eta)-{{h(\omega_{l+1})}\over{h(\omega_{k})}}B_{k,l}(\la_0)C_{k-1,l+1}(\la_0+2\eta)\nonumber\\
&=&\prod_{i=1}^N{{h(\la_{0}-\la_i)}\over{h(\la_{0}-\la_i+2\eta)}}
\eeqa
As the operator $D_{l}$ is diagonal it can be inverted easily, and the operator $C_{l}$ possesses a global vacuum which is annihilated by its action. This enables one to compute $A_{l}$
\beqa
{\tilde A}_{l}(\la_0)&=&{{h(\omega_{{{k+l-N}\over{2}}})}\over{h(\omega_{k})}}
 \left\{
\otimes_{i=1}^N
{\left(\matrix{
1&0\quad\quad\cr
0&b(\la_i-\la_0)^{-1}\cr
}\right)}_{[i]}\right.\nonumber\\&&\nonumber\\
&+&\sum_{i=1}^N {{c_{k-1}(\la_0-\la_i)
c'_{l}(\la_0-\la_i)}
\over{b(\la_0-\la_i)}
}
{\left(\matrix{
0&0\cr
0&1\cr
}\right)}_{[i]}
\otimes_{j\ne i}
{\left(\matrix{
{{b(\la_0-\la_j)}\over{b(\la_i-\la_j)}}&0\quad\cr
0&b(\la_j-\la_i)^{-1}\cr
}\right)}_{[j]}\nonumber\\&&\nonumber\\
&+&\left.\sum_{i\ne j}^N
{
{c_{k-1}(\la_0-\la_i){c}'_{l}(\la_0-\la_j)}\over
{b(\la_j-\la_k)}}\,
\sigma_-^i\otimes\sigma_+^j\otimes_{k\neq i,j}
{\left(\matrix{
{{b(\la_0-\la_k)}\over{b(\la_j-\la_k)}}&\quad\quad0\cr
0\quad\quad&{b(\la_k-\la_i)}^{-1}\cr
}\right)}_{[k]} \right\}
\label{A}
\eeqa
\section{Conclusion}
The form of the $F$-matrix, Eq. (\ref{Fmatrix}) and the appearance of the monodromy matrix in the basis supplied by the $F$-matrix, Eq's (\ref{ops}), (\ref{A}), are completely analogous to what has been found in \cite{ms} and \cite{abfr} for the rational and trigonometric models. The concrete expressions for ${\tilde{A}},\,{\tilde{B}},\,{\tilde{C}},\,{\tilde{D}}$ are in particular manifestly symmetric with respect to exchanges of the local inhomogeneity parameters $\la_i$. The quasiparticle operators ${\tilde{B}}$ and ${\tilde{C}}$ are free from polarization effects due to non-local exchange terms.\\
The argument used in \cite{abfr} - borrowed from \cite{ms} - concerning the identification of operators corresponding to different entries of the monodromy matrix relied on the $sl(n)$ symmetry of the rational model. It is not available for the trigonometric and elliptic model. The recursive procedure followed instead in the preceding section is equally applicable to the $XXX$, $XXZ$ and $XYZ$ model.\\
It seems rather plausible in view of the formal similarities of the rational, trigonometric and elliptic models that some version of Sklyanin's functional Bethe ansatz should also be feasible in the latter case as has already been achieved for the $XYZ$ Gaudin magnet in \cite{sklytak}.\\\vspace*{1.5cm}\\
{\bf{Acknowledgement:}} We thank Frank G\"ohmann for a discussion. H.B. and R.H.P. acknowledge support of  the Alexander von Humboldt Foundation. R.F. was supported by the TMR network contract FMRX-CT96-0012 of the European Commission. 
\begin{appendix}
\setcounter{secnumdepth}{-1}
\section{Appendix}
In this appendix we derive the identity used in the recursion relation (\ref{recrel}) using the factorizing $F$-matrix for $N$ sites (for the sake of simplicity we omit the explicit dependence on $l$, which can be restored by using  Eq.(\ref{complaw})):
\beqa
F_{1\ldots N}&=& \sum_{{{\alpha}}\in {\mathbb{Z}}_2^N}
P_{{\alpha}} R_{1\ldots N}^{\sigma_{{\alpha}}}(z_{1},\ldots,z_{N})\nonumber\\
P_{{\alpha}}&=&\prod_{i=1}^N P_{i}^{\alpha_i}\label{F1}
\eeqa
and $P_{i}^{\alpha_i}$ projects on the ${\alpha_i}$-th component in the $i$-th space.\\
With a given ${\alpha}$ the permutation $\sigma_{\alpha}\in{\cal{S}}_N$ is uniquely determined by the following algorithm: \\
Let $k$ be the number of 1's in the sequence $\alpha_1,\ldots ,\alpha_N$. For $i\leq k$ we put $\sigma_{\alpha}(i)=l$, where $l$ is the label of the $i$-th 1 in the above mentioned sequence.\\
For $i>k$ we put $\sigma_{\alpha}(i)=k+l'$, where $l'$ is the label of the $i-k$-th 2 in the sequence.\\
The permutation constructed this way is the only one satisfying the constraints 
\beqa
\alpha_{\sigma_{\alpha}(i+1)}&\geq&\alpha_{\sigma_{\alpha}(i)}\quad \mbox{if}\quad\sigma_{\alpha}(i+1)>\sigma_{\alpha}(i)\nonumber\\
\alpha_{\sigma_{\alpha}(i+1)}&>&\alpha_{\sigma_{\alpha}(i)}\quad \mbox{if}\quad\sigma_{\alpha}(i+1)<\sigma_{\alpha}(i)\;.
\label{cond}
\eeqa
Now we will prove that $$F_{12\ldots N}=F_{23\ldots N}\left(P_1^1+P_1^2 T_{1,23\ldots N}\right)\;.$$
If $\alpha \in {\mathbb{Z}}^{N}_2$ has the form $\left(1,\alpha_2,\ldots,\alpha_N\right)\equiv\left(1,\tilde{\alpha}\right);\;\tilde{\alpha}=\left(\alpha_2,\ldots,\alpha_N\right)\in{\mathbb{Z}}^{N-1}_2$, we have $P_{{\alpha}}=P_1^1P_{{\tilde{\alpha}}}$ and $\sigma_{\alpha}=\sigma_{\tilde{\alpha}}$ (here and in what follows we identify the symmetric group ${\cal{S}}_{N-1}$ acting on the elements $(2,\ldots,N)$ as the subgroup of ${\cal{S}}_{N}$ acting on the elements $(1,2,\ldots,N)$). Thus the part of the sum of (\ref{F1}) corresponding to such $\alpha$'s exactly gives $P_1^1 F_{2\ldots N}$.\\
Now consider the case when $\alpha=\left(2,\alpha_2,\ldots,\alpha_N\right)\equiv\left(2,\tilde{\alpha}\right)$. It is easy to check that $P_{{\alpha}}=P_1^2P_{{\tilde{\alpha}}}$ and $\sigma_{\alpha}=\sigma_{\tilde{\alpha}}\sigma_{1,\tilde{2}}\ldots\sigma_{1,\widetilde{k+1}}$ where $k$ is the number of 1's in $\tilde{\alpha}$ and $\sigma_{i,j}$ are elementary transpositions and $\tilde{i}=\sigma_{\tilde{\alpha}}(i)$. We have
\beqa
P_1^2P_{\tilde{\alpha}}R^{\sigma_{\tilde{\alpha}}}T_{1,23\ldots N}&=&P_{{\alpha}}T_{1,\sigma_{\tilde{\alpha}}(2)\ldots \sigma_{\tilde{\alpha}}(N)}R^{\sigma_{\tilde{\alpha}}}\nonumber\\
&=&P_{{\alpha}}R_{1,\sigma_{\tilde{\alpha}}(k+1)}\ldots R_{1,\sigma_{\tilde{\alpha}}(2)}R^{\sigma_{\tilde{\alpha}}}\nonumber\\
&=&P_{{\alpha}}R^{\sigma_{\tilde{\alpha}}\sigma_{1,{\tilde{2}}}\ldots\sigma_{1,{\widetilde{k+1}}}}=P_{{\alpha}}R^{\sigma_{\alpha}}
\eeqa
where we have taken into account that ${\tilde{\alpha}}_{\widetilde{k+2}}=\ldots ={\tilde{\alpha}}_{\tilde{N}}=2$ and that $P_1^2P_i^2R_{1i}=P_1^2P_i^2\id$.
\end{appendix}

\vfill\eject
\end{document}